\documentclass[%
 aip,
 jmp,%
 amsmath,amssymb,
reprint,%
]{revtex4-1}

\usepackage{graphicx}
\usepackage{dcolumn}
\usepackage{bm}
\usepackage{hyphenat}
\usepackage{subfigure}

\begin{document}

\preprint{AIP/123-QED}

\title{Highly reliable low-energy writing of bits in non-volatile multiferroic memory based on 180-degree magnetization switching with voltage-generated stress}

\author{Ayan K. Biswas}

\author{Supriyo Bandyopadhyay}
\affiliation{%
Department of Electrical and Computer Engineering, Virginia Commonwealth University, Richmond, Virginia 23284, USA
}
%

\author{Jayasimha Atulasimha}%
\affiliation{
Department of Mechanical and Nuclear Engineering, Virginia Commonwealth University, Richmond, Virginia 23284, USA
}%

\date{\today}

\begin{abstract}

Rotating the magnetization of a magnetostrictive
nanomagnet with electrically generated mechanical strain dissipates miniscule amount of energy compared to any other rotation method and would have been the ideal
method to write bits in non-volatile magnetic memory, except strain cannot ordinarily rotate the magnetization of magnet by
more than 90$^{\circ}$ and ``flip'' it. Here, we describe a scheme to achieve complete 180$^{\circ}$ rotation of the magnetization of a nanomagnet with strain
that will enable writing of binary bits in non-volatile magnetic memory implemented with magneto-tunneling junctions whose soft layers are two-phase magnetostrictive/piezoelectric
multiferroics. At room temperature, this writing method results in:
(1) energy dissipation $<$ 6200 kT per bit,
(2) write error probability $<$ 10$^{-6}$, (3) write time of $\sim$ 1 ns,
and (4) low read error.  This could potentially lead to a new genre of non-volatile memory that is
extremely reliable, fast and, at the same time, ultra-energy-efficient.

\end{abstract}

\maketitle

Magnetic random access memory (MRAM) is typically implemented with a magneto-tunneling junction (MTJ) comprising a hard
and a soft
ferromagnetic layer separated by a spacer that acts as a tunnel barrier. The soft layer is
shaped like an elliptical disk that has two stable (mutually anti-parallel)
magnetization  states along its major axis. The hard layer is permanently magnetized parallel
to one of those states.
When the soft
layer's magnetization is parallel to that of the hard layer, the MTJ's resistance is low and encodes one
binary bit (say `0'), and when it is anti-parallel, the MTJ's resistance is high and encodes the other bit (say `1').
Writing a bit involves orienting the magnetization of the soft layer either parallel or anti-parallel
to that of the hard layer.

The oldest bit writing scheme used a magnetic field to flip the soft layer's magnetization and that field was generated
with an on-chip current. More recent schemes flip the magnetization with
a spin transfer torque (STT) generated by passing a current through the MTJ \cite{Ralph2008}, or domain wall motion
induced by the current \cite{Yamanouchi2004}, or by manipulating Rashba spin-orbit interaction
at interfaces \cite{Miron2011}. These schemes are extremely dissipative and result in dissipating
$\sim$10$^7$ kT of energy per bit at room temperature \cite{KangWang2013}. A more energy-efficient scheme is to rotate
the magnetization with a current utilizing the spin-Hall effect \cite{Liu2012} or spin-orbit torque in
magnetically doped topological insulators \cite{Wang2014},  or use voltage-generated
uniaxial strain/stress in a shape anisotropic magnetostrictive-piezoelectric (multiferroic) nanomagnet
\cite{Atulasimha2010,Roy2011,Mohammed2011,Roy2013}. The strain/stress is generated by applying
an electrical voltage across
the piezoelectric layer of the multiferroic nanomagnet which transfers the resulting strain to
the magnetostrictive layer and rotates its magnetization.

Unfortunately, strain/stress can rotate the magnetization of a nanomagnet by only up to
$\sim$90$^{\circ}$, which means that it is not able to ``flip'' the magnetization since that
requires a $\sim$180$^{\circ}$ rotation. Once the stress is withdrawn,
the magnetization, which has rotated by 90$^{\circ}$ and is now in an unstable state, will have roughly equal likelihood of returning to the
original stable orientation (not flipping, or 0$^{\circ}$ rotation) or flipping to the other stable orientation
(180$^{\circ}$ rotation). That makes the
flipping only $\sim$50\% likely, which is untenable. However, if the stress is
withdrawn {\it as soon as} the magnetization vector has rotated by 90$^{\circ}$ from the original orientation, then
a residual torque due to the magnetization vector's out of plane component may continue to rotate
it beyond 90$^{\circ}$ and achieve a ``flip" with very high probability ($>$ 99.99\% at room temperature) \cite{Roy2013}.
Such precise withdrawal requires a feedback mechanism that determines when the magnetization has completed
the 90$^{\circ}$ rotation and feeds that information back to the voltage generator
to withdraw the stress at exactly the right juncture \cite{Roy2013}. The need for such feedback circuitry makes this strategy
unappealing since it introduces additional energy dissipation and complexity.

A clever idea to circumvent this problem
is
to apply a permanent magnetic field
along the minor axis of the elliptical soft layer that dislodges  the stable magnetization orientations
from the major axis and places them along two axes that are in the plane of the magnet and mutually perpendicular. The
hard layer of the MTJ is now
magnetized parallel to one of these axes. If the soft layer has positive
magnetostriction, then
applying uniaxial tensile stress along one of these two axes brings the magnetization to the stable state along that axis,
while applying compressive stress takes it to the other stable state (the
situation is opposite if the magnetostriction coefficient is negative) \cite{Tiercelin2011,Giordano2012,Giordano2013}.
Therefore, one type of stress (say, compressive) makes the hard and soft layer's magnetization parallel and the
other type (say, tensile) makes them mutually perpendicular. Thus, by choosing the {\it sign} of the stress, we can
write either bit `0' or bit `1', regardless of the initially stored bit.
The advantages are: 1) the stress withdrawal timing is not critical and no feedback mechanism is needed,
2) the final state is always stable; hence, the magnetization remains in the final state after the stress is withdrawn
making the error probability very low ($<$ 10$^{-6}$ at room temperature), and 3) by choosing the sign of the stress (compressive or tensile), one can
deterministically write either bit `0' or bit `1', without knowing what the initial stored bit was (i.e. there
is no need to read the stored bit before re-writing it). On the flip side, the disadvantages are: 1) An external bias magnetic field
is needed to dislodge the
stable magnetization orientations from the ellipse's major axis, and 2) the separation angle
between the stable magnetization orientations is now $\Theta \approx 90^{\circ}$ which makes it
harder to distinguish bit `1' from bit `0' when the latter are read via the resistance of the MTJ.
The ratio of the MTJ resistances corresponding to bit `1' and bit `0' is
$\left (1 + \eta_1 \eta_2 \right )/\left (1 + \eta_1 \eta_2 cos \Theta \right )$ when the magnetization of the
hard layer is aligned along the stable direction representing the bit `0' \cite{book}. Here, $\eta_1$ and $\eta_2$ are the
spin injection/detection efficiencies at the two ferromagnet/spacer interfaces of the MTJ. The resistance
ratio is largest when $\Theta = 180^{\circ}$ and smaller when $\Theta = 90^{\circ}$. In fact, if $\eta_1 =
\eta_2 = 1$, then the resistance ratio is infinite when $\Theta = 180^{\circ}$ and only 2:1 when
$\Theta = 90^{\circ}$. Therefore, it is imperative to increase $\Theta$ and bring it as close to
180$^{\circ}$ as possible. We were recently able to increase
$\Theta$ to 132$^{\circ}$ by applying stress (of the same sign) along {\it two different} directions (instead of
applying compressive and tensile stress along the same direction)
to write the two bits \cite{Biswas2014}. However, this improves the resistance ratio only moderately.

In this paper, we propose a scheme that can increase $\Theta$ to 180$^{\circ}$ and also eliminates the bias
magnetic field needed in Refs. [\onlinecite{Tiercelin2011,Giordano2012,Giordano2013,Biswas2014}]. The only penalty
we pay is that the write cycle must be preceded by a read cycle, i.e. we cannot deterministically
write either bit `0' or bit `1' without first knowing what the initial stored bit was. Since reading is
both faster than writing and dissipates far less energy, this penalty is minor.
Figure 1(a) shows the
schematic design of the memory element where the bit storing MTJ is placed on top
of a piezoelectric Lead Zirconate Titanate (PZT) thin film deposited on an n$^+$-Silicon substrate. The major
axes of the elliptical hard and soft layers are collinear and the hard layer is permanently magnetized in one direction
along its major axis. Two pairs of electrode
pads are delineated on the PZT film such that the line joining one pair subtends an angle of 30$^{\circ}$ with the
common major
axis
of the two layers and the other pair subtends an angle of 150$^{\circ}$. The magnetostrictive elliptical soft layer is
in elastic contact with the PZT thin film and has a major axis $a$ = 110 nm, minor axis $b$ = 90 nm and thickness $d$
 = 6 nm. These
dimensions ensure that the soft nanomagnetic layer has a single domain \cite{Cowburn1999} and the in-plane potential
energy barrier
separating the stable magnetization orientations along its major axis is $\sim$62.5 kT at room temperature. The probability of spontaneous
magnetization
flipping between the two stable states due to thermal noise (static error probability) is therefore
$\sim e^{-62.5}$ per attempt
\cite{Brown1963}, leading to memory retention time $(1/f_o)e^{62.5} = 4.4\times 10^7$ years, assuming the attempt
frequency $f_o$ is 1 THz \cite{Gaunt1977}.

One pair of electrodes has edge dimension of 120 nm and the other has edge
dimension of 80 nm whereas the thickness of the PZT thin film is 100 nm. These dimensions are needed to ensure the
following:
(1) the line joining the centers of each pair of pads subtends either 30$^{\circ}$ or 150$^{\circ}$ with
the common major axis of the ellipses, (2)
the spacing between the facing edges of the pads in either pair is comparable to the pads' edge dimension and also the
PZT film thickness, and (3) no two pads overlap. The hard layer of the MTJ is implemented with a synthetic
anti-ferromagnet (SAF). When the magnetizations of the soft and hard
layer are parallel (state $\Psi_0$ in Figure 1(b)), the stored bit is `0', and when they are anti-parallel
(state $\Psi_1$), the stored bit is `1'.

In order to write, say, bit `1', we first read the resistance of the MTJ to determine what the stored
bit is. If it is bit `1', we do nothing. If it is bit `0', then the magnetization of soft layer is at $\Psi_0$ and
we must switch it to $\Psi_1$
(Figure 1(b)). To accomplish this, we apply a voltage between the electrode pair AA$^{\prime}$ and the grounded
n$^+$-Silicon substrate.

\begin{figure}[!ht]%
 \centering
 \subfigure[Device schematic]{
  \includegraphics[width=3.4in]{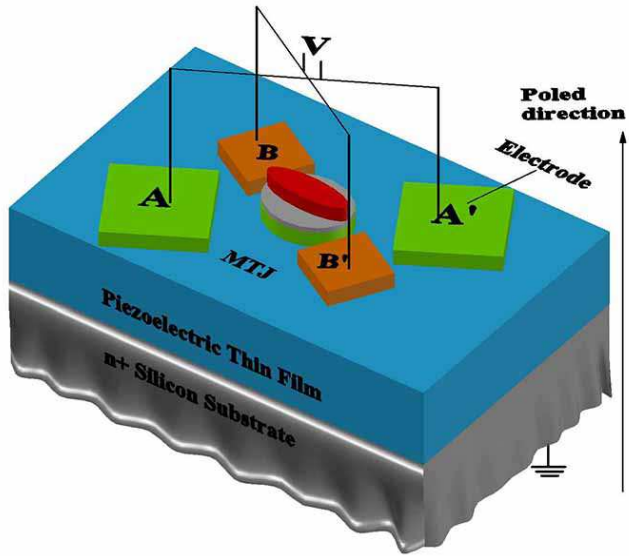}
   \label{fig:fig1(a)}}
 \subfigure[2-dimensional view of the device]{
  \includegraphics[width=3.4in]{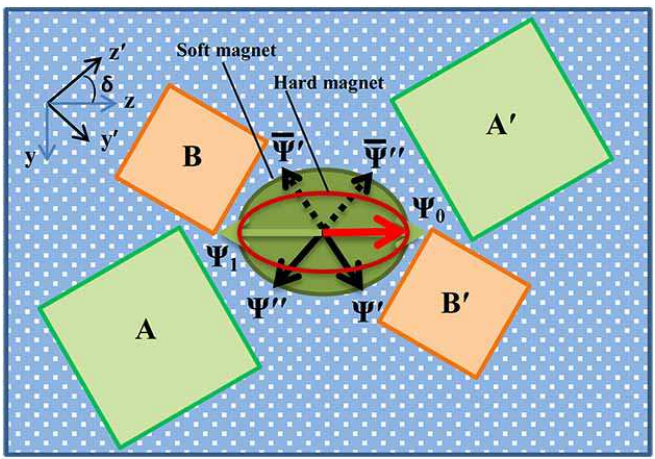}
   \label{fig:fig(b)}}%
 \caption[Caption of figure]{{\bf Schematic of memory element}. (a) The PZT film has a thickness of $\sim$100 nm and is
deposited on a conducting n$^+$-Si substrate. It is poled with an electric field in the direction shown. The ratio
of the distance
between the facing edges of the electrodes to the electrode lateral dimensions is 1.67. (b) The fixed
magnetization orientation of the top (hard) magnet is denoted by the red arrow, and the two stable magnetization orientations
of the bottom (soft) magnet are denoted by green arrows. The MTJ resistance is high when the soft magnet's magnetization
is aligned along $\Psi_1$ and the resistance is low when the soft magnet's magnetization is aligned along $\Psi_0$.
Also shown are the orientations of the intermediate states $\Psi^{\prime}$, $\Psi^{\prime \prime}$,
$\overline{\Psi}^{\prime}$, $\overline{\Psi}^{\prime \prime}$. The eccentricity of the hard magnet is more than that
of the soft magnet which helps to make the hard magnet ``hard'' and the soft magnet ``soft''.}
  \label{fig:Fig1}
\end{figure}

Since the electrode in-plane dimensions are comparable to the piezoelectric film thickness, the out-of-plane ($d_{33}$)
expansion/contraction and the in-plane ($d_{31}$) contraction/expansion of the piezoelectric regions underneath the
electrodes
produce a highly localized strain field under the electrodes \cite{Lynch2013}. Furthermore, since the electrodes are separated by a
distance 1-2 times the PZT film thickness, the interaction between the local strain fields below the electrodes will lead
to a biaxial strain in the PZT layer underneath the soft magnetic layer \cite{Lynch2013}. This biaxial strain (compression/tension
along the line joining the electrodes and tension/compression along the perpendicular axis) is transferred to the soft
magnetostrictive magnet by elastic coupling. If the magnetostriction coefficient of the latter is positive, then voltage of the correct
polarity applied between AA$^{\prime}$ and ground will generate compressive stress in the soft magnet,
rotating its magnetization, while if the magnetostriction coefficient is negative, then voltage of the
opposite polarity will cause the rotation. This rotation happens despite any substrate clamping and despite the fact that the electric field
in the PZT layer just below the magnet is approximately zero since the metallic magnet shorts out the field \cite{Lynch2013}.

We will assume that the magnetostriction coefficient
of the soft layer is positive and the applied voltage between AA$^{\prime}$ and ground (of the right polarity)
has  generated compressive stress along the line AA$^{\prime}$
and tensile stress in the direction perpendicular to this line. In that case, the magnetization
will rotate away from $\Psi_0$ towards $\Psi_1$. Once steady state is reached
and the magnetization settles at some intermediate state $\Psi^{\prime}$ which is roughly perpendicular to
the axis joining the electrodes AA$^{\prime}$, the voltage
at
AA$^{\prime}$ is withdrawn (timing is not critical) and a voltage is applied between BB$^{\prime}$ and the grounded
substrate that will
rotate the
magnetization farther towards $\Psi_1$. Finally, upon reaching the new steady-state at
$\Psi^{\prime \prime}$ which is roughly perpendicular to the line joining the electrode pair BB$^{\prime}$,
the voltage at BB$^{\prime}$ is withdrawn (again, timing is not critical) and the magnetization vector
will rotate spontaneously to the closer of the two global energy minima, which is
$\Psi_1$. This results in flipping the magnetization and writing the desired bit `1'. Writing bit `0', when
the initial stored bit was `1', is exactly
equivalent and hence not discussed. Note that a two-phase clock is required to flip the bit -- one phase tied to
AA$^{\prime}$ and the other to BB$^{\prime}$.

In order to estimate the energy dissipated in writing the bit, the minimum time required to write, and the write
error probability, we have carried out stochastic Landau-Lifshitz-Gilbert calculations in the manner of
Ref. [\onlinecite{Biswas2014}]. For the sake of simplicity, we always consider uniaxial strain along the line joining
the two electrodes of a pair, but the strain is actually biaxial resulting in tension/compression along that line and
compression/tension along the perpendicular direction. The torques on the magnetization
vector due to these two components {\it add}. Therefore,
we {\it underestimate} the torque that makes the magnetization vector rotate, which makes all our dissipation,
error probability and
switching delay figures {\it conservative}.

\begin{figure}[ht]
\begin{minipage}[b]{0.48\linewidth}
\centering
\includegraphics[width=\textwidth]{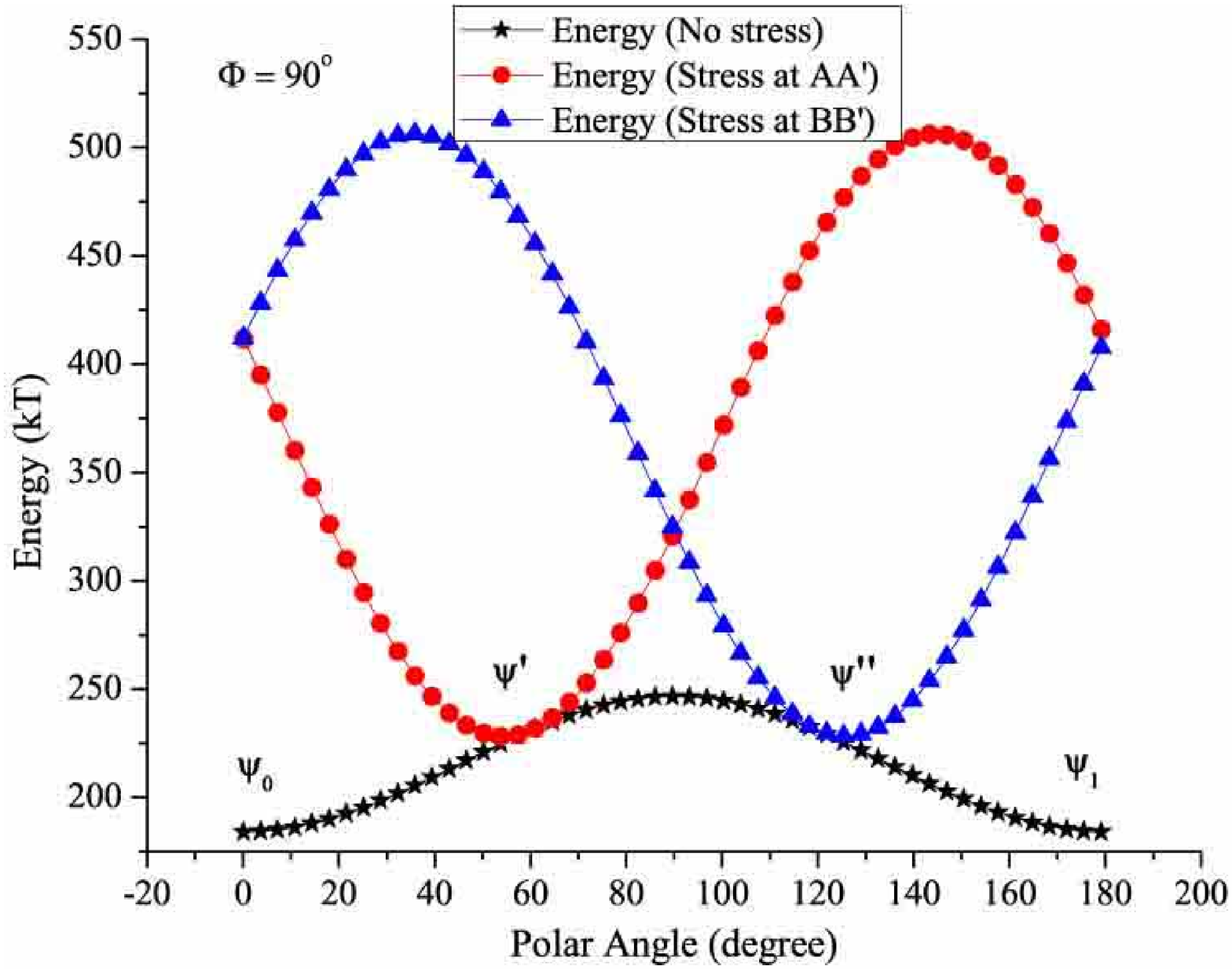}
\label{fig:figure2a}
\end{minipage}
\hspace{0.2cm}
\begin{minipage}[b]{0.48\linewidth}
\centering
\includegraphics[width=\textwidth]{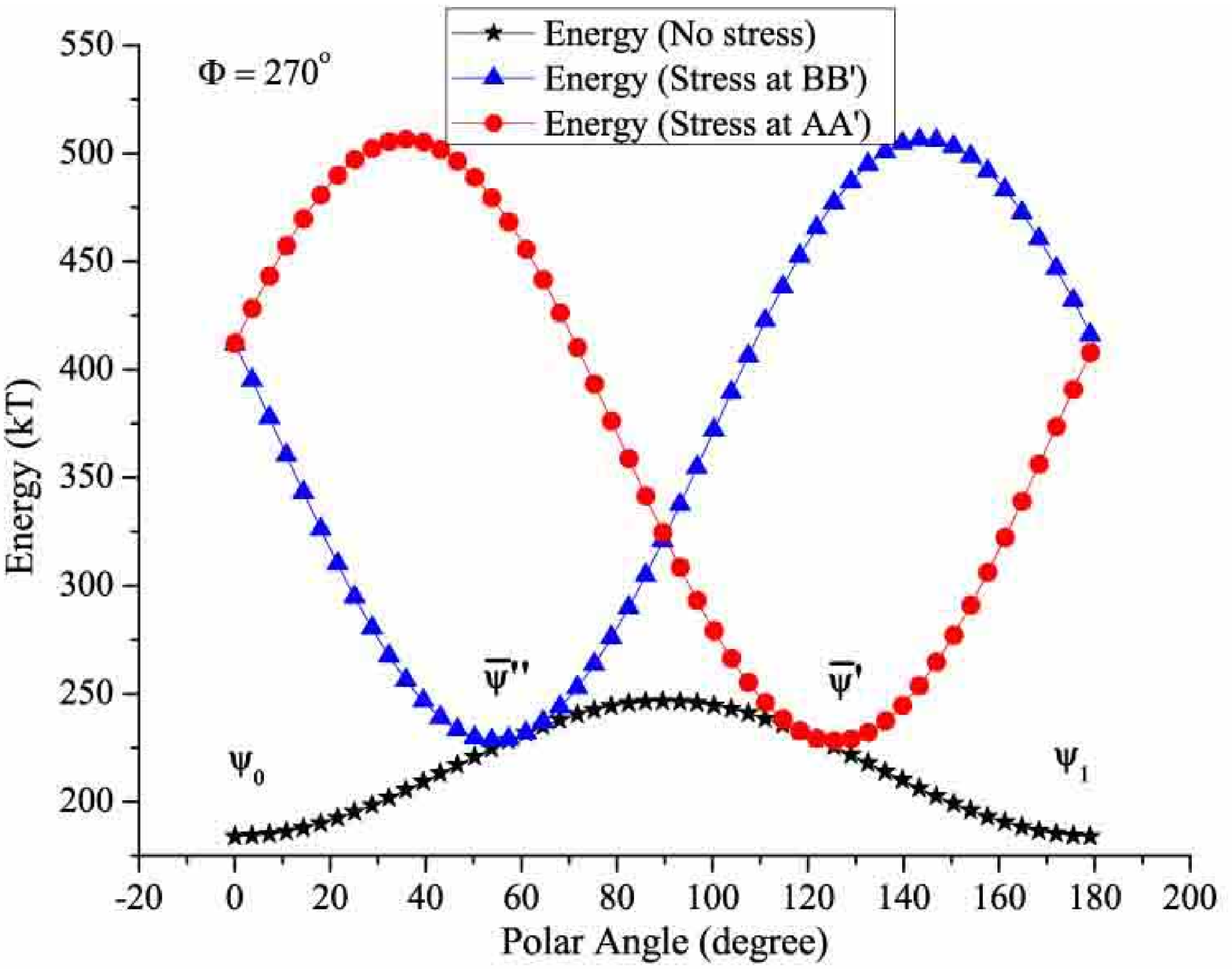}
\label{fig:figure2b}
\end{minipage}
\caption{\label{fig:fig2} Potential energy profiles of a Terfenol-D magnetostrictive nanomagnet of stated
dimensions when the
magnetization vector is constrained to the plane of the magnet ($\phi$ = 90$^{\circ}$,
270$^{\circ}$). The three curves show the profiles
when no electrode pair is activated, electrode pair AA$^{\prime}$ is activated and
electrode pair BB$^{\prime}$ is activated. Activating electrode pair AA$^{\prime}$ creates global energy minima at $\Psi^{\prime}$
($\phi = 90^{\circ}$) and $\overline{\Psi}^{\prime}$ ($\phi = 270^{\circ}$), whereas
activating pair BB$^{\prime}$, creates global minima at $\Psi^{\prime\prime}$ ($\phi = 90^{\circ}$)  and
$\overline{\Psi}^{\prime \prime}$ ($\phi = 270^{\circ}$).}
\end{figure}

Figure \ref{fig:fig2} shows the potential energy profile of the nanomagnet in the magnet's plane ($\phi$ = 90$^{\circ}$,
270$^{\circ}$)
as a function of the polar
angle $\theta$ subtended by the magnetization vector with the common major axis of the elliptical hard and soft layers (z-axis). The three profiles
correspond to the situations when neither electrode pair is activated, electrode pair AA$^{\prime}$ is activated, and
electrode pair BB$^{\prime}$ is activated.

Consider the case when the magnetization of the nanomagnet is
initially
in the stable state $\Psi_{0}$ (initial stored bit is `0'). If the electrode pair AA$^{\prime}$
is activated, a compressive uniaxial stress component
is generated along the line joining that electrode pair, which will
rotate the magnetization vector to $\Psi^{\prime}$ since that corresponds to the
only accessible global energy minimum (see the energy profile corresponding to $\phi = 90^{\circ}$ in Fig. 2).
The other global minimum at $\overline{\Psi}^{\prime}$ is inaccessible owing to the energy barrier between $\Psi_0$ and
$\overline{\Psi}^{\prime}$
(see energy profile corresponding to $\phi = 270^{\circ}$ in Fig. 2; the peak of the energy barrier
separating $\Psi_0$ and
$\overline{\Psi}^{\prime}$ is located roughly at $\theta = 35^{\circ}$). In other words, the
magnetization will rotate clockwise instead of anti-clockwise in Fig. 1(b).
Next, de-activating AA$^{\prime}$
and activating BB$^{\prime}$ causes a uniaxial compressive stress component along the
line joining BB$^{\prime}$ that will rotate the magnetization clockwise to the new global
energy minimum $\Psi^{\prime \prime}$, which is the only accessible one. Finally, removal of stress
will drive the magnetization to $\Psi_{1}$ (writing the new bit `1') since it is
the only accessible global energy minimum at that point. The other global energy minimum at $\Psi_{0}$ is
inaccessible because
of the energy barrier
between $\Psi^{\prime \prime}$ and $\Psi_{0}$. The height of this energy barrier $>$ 20 kT which prevents
the magnetization from migrating to $\Psi_{0}$ as opposed to $\Psi_{1}$.

If we activate the electrode pairs in opposite sequence, i.e. BB$^{\prime}$ first and then AA$^{\prime}$, the magnetization
will first rotate anti-clockwise from $\Psi_{0}$ to $\overline{\Psi}^{\prime \prime}$, then
anti-clockwise to $\overline{\Psi}^{\prime}$ and finally anti-clockwise
to $\Psi_{1}$ (see the energy profile corresponding to $\phi = 270^{\circ}$). Therefore, the sequence does not matter;
activating the electrodes in either sequence always flips the bit - either by clockwise rotation or anti-clockwise
rotation depending on the sequence. It is easy to verify that the same is true if the
initial stored bit was `1' instead of `0'.

\begin{figure}[ht]
\begin{minipage}[b]{0.49\linewidth}
\centering
\includegraphics[width=\textwidth]{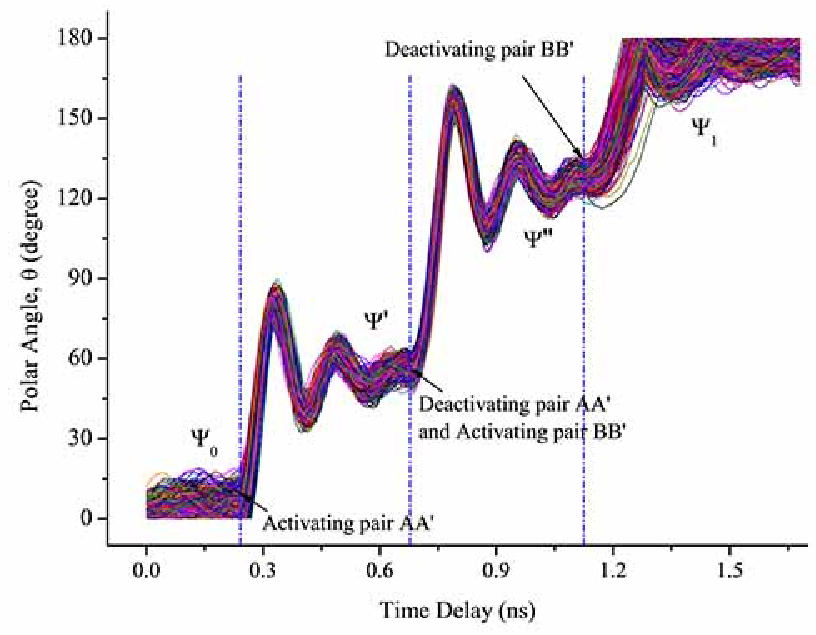}
\label{fig:figure3a}
\end{minipage}
\begin{minipage}[b]{0.49\linewidth}
\centering
\includegraphics[width=\textwidth]{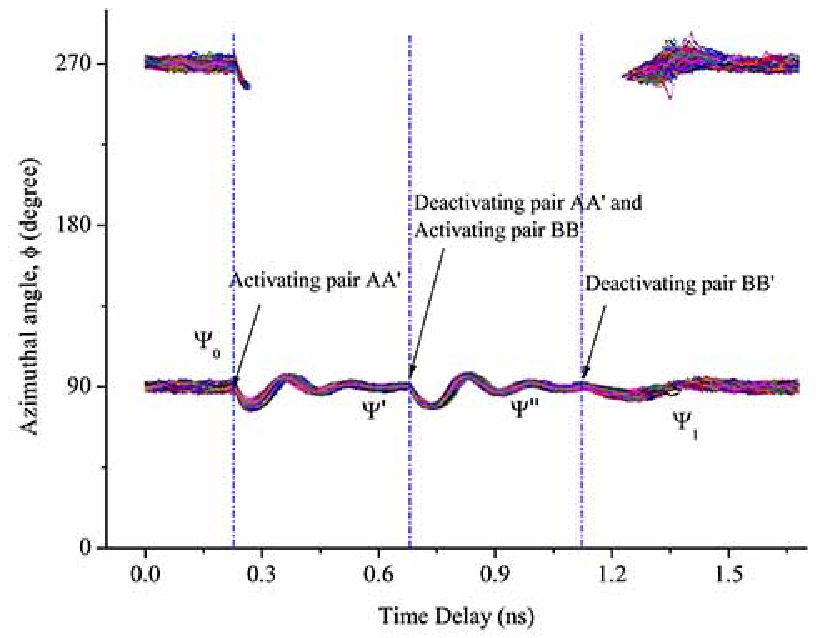}
\label{fig:figure3b}
\end{minipage}
\caption{\label{fig:fig3} Magnetization dynamics at room temperature.
Polar angles (left) and azimuthal angles (right) of randomly chosen 1,000 trajectories out of
10$^6$ trajectories plotted as a function of time. The trajectories are all slightly different from each other
because of random thermal noise included in the simulation used to generate these trajectories in the manner of ref. [15].
The instants at which the electrode pairs are activated and deactivated are shown.}
\end{figure}

In Fig. \ref{fig:fig3}, we show 1,000 randomly chosen switching trajectories (magnetization orientation
$\theta, \phi$ versus time) out of 10$^6$ trajectories simulated in the presence of room temperature thermal noise.
These switching trajectories are generated from stochastic Landau-Lifshitz-Gilbert simulations in the
manner of ref. [\onlinecite{Biswas2014}]. The initial values of $\theta$ and $\phi$ are chosen from their
thermal distributions around $\theta = 0^{\circ}$ with appropriate weight \cite{Biswas2014}, the fluctuation of
the magnetization around the initial orientation is simulated for 2.3 ns and then the pair AA$^{\prime}$ is activated
(stress is turned on).
The intermediate steady state $\Psi^{\prime}$ ($\theta = 60^{\circ} \pm 4^{\circ}$) is reached by all 10$^6$ trajectories
within 0.45 ns after activation, at which point  AA$^{\prime}$ is deactivated and BB$^{\prime}$ is activated. The next
intermediate steady state $\Psi^{\prime \prime}$ ($\theta = 120^{\circ} \pm 4^{\circ}$) is reached by all trajectories
within another 0.45 ns, and then BB$^{\prime}$ is deactivated (stress is turned off).
The simulation of the trajectories is continued until for every one of 10$^6$ trajectories,
$\theta$ reaches within 4$^{\circ}$ of
180$^{\circ}$ (successful flip). This takes another 0.46 ns.  The trajectories in Fig. \ref{fig:fig3} are all slightly different from each other since
they are probabilistic in the presence of room temperature thermal noise. One million trajectories were simulated
and all of them showed successful transition from $\theta \approx
0^{\circ}$ to $\theta \approx 180^{\circ}$, implying that the switching failure probability is $< 10^{-6}$.
Therefore, the minimum switching delay for $< 10^{-6}$ error probability
is 0.45 + 0.45 + 0.46 ns = 1.36 ns. This is the minimum time needed for {\it all} 10$^6$ switching trajectories
to complete flipping.

It is obvious that the present scheme has the shortcoming that it will erroneously write the
wrong bit every time the stored bit happens to be the desired bit (since the stored bit is
always flipped in the write step). Therefore, a write cycle must be {\it preceded} by a read cycle to determine the stored
bit.
If the stored bit is the same as the desired bit, no action is taken. Otherwise, the bit is flipped
following the above procedure. This requires an extra read cycle, but it also saves time and
energy by obviating the write cycle whenever the stored and desired bits are the same. Since writing is
both slower and more dissipative than reading, there may be an overall gain.

The write error probability can be reduced to zero by writing the bit, then reading it to verify if it
was written correctly, re-writing it if it was written incorrectly, followed by another read and so on, until
the bit is verified to have been written correctly. Alternately, we can always carry out a
fixed number of write/verification cycles. The error probability after $n$ such cycles is $10^{-6n}$ since
it is the probability of having written the bit incorrectly $n$ times in a row. Because it will
be an overkill to reduce the write error
probability to below the static error probability of $e^{-62.5} = 10^{-27}$, just four ($n$ = 4)
read/verification cycles will be sufficient. However, this increases the write time. Even if the
bit was written correctly in the first attempt, we will still need three additional idle cycles
since all bits are written simultaneously in parallel. This will increase the effective write time
to 1.36$\times$4 ns = 5.44 ns (again assuming that the read time is negligible compared to the write time),
resulting in a clock rate of 180 MHz.

The results in Fig. \ref{fig:fig3} were generated assuming the following material parameters for the magnet (Terfenol-D):
saturation magnetization $M_s = 8\times 10^5$ A/m,
magnetostriction coefficient $(3/2)\lambda_s = 90 \times 10^{-5}$, Young's modulus $Y$ = 80 GPa, and Gilbert damping
coefficient $\alpha = 0.1$ \cite{Abbundi1977,Ried1998,Kellogg2008}. We also assume: strain $\epsilon(t) =
3.75 \times 10^{-4} $ (stress = 30 MPa).

In ref. [\onlinecite{Lynch2013}], the electric field needed to generate a local strain of $\sim$10$^{-3}$ in the magnet
was 3 MV/m. Using a linear interpolation, the electric field needed to generate a strain of $3.75\times 10^{-4}$ would be
1.125 MV/m. Therefore, the potential that needs to be applied to the electrodes is 1.125 MV/m $\times$ 100 nm = 112.5 mV.

The energy dissipated in writing the bit has two components: (1) the {\it internal} dissipation in the nanomagnet due to
Gilbert damping,
which is calculated in the manner of Ref. [\onlinecite{Roy2012}] for each trajectory (the mean dissipation is the
dissipation averaged over all trajectories that result in correct switching); and (2) the {\it external} (1/2)$CV^2$
dissipation associated with applying the voltage between the electrodes and the grounded substrate which act as a
capacitor. Since the piezoelectric response of PZT is much faster than the magnet switching \cite{Ramesh2004}, we can
view the strain generation as instantaneous.

The larger electrodes have a lateral dimension of 120 nm and the PZT film thickness is 100 nm. Therefore, the
associated capacitance is $C$ = 1.275 fF, if we assume that the relative dielectric constant of PZT is 1000. Since the two
electrodes of a pair are always activated together, the external energy dissipation will be twice (1/2)$CV^2$ dissipation
and that value is 3896 kT at room temperature ($V$ = 112.5 mV). The smaller electrode pair has a lateral
dimension of 80 nm and hence a smaller capacitance of 0.567 fF.
Hence, it dissipates $CV^2$ energy of 1733 kT. The mean internal dissipation could depend on whether the initial stored bit was `0' or
`1', and we will take the higher value.
In this case, the higher value was 514 kT, thus making the total dissipation 6143 kT which is at least two orders of
magnitude
less than what spin-transfer-torque memory STT-RAM dissipates in a write cycle \cite{KangWang2013}.

In conclusion, we have provided a straintronic bit writing scheme that results in low energy dissipation, low write
and read error rate, and fast writing speed. The only disadvantage is that every write cycle must be preceded by a
read cycle, but it is a minor penalty.

This work was supported by the US National Science Foundation under grants ECCS-1124714 and CCF-1216614. J. A.
would also like to acknowledge the NSF CAREER grant CCF-1253370.

\providecommand{\latin}[1]{#1}
\providecommand*\mcitethebibliography{\thebibliography}
\csname @ifundefined\endcsname{endmcitethebibliography}
  {\let\endmcitethebibliography\endthebibliography}{}

\end{document}